# Approaches for Uncertainty Quantification of AI-predicted Material Properties: A Comparison

Francesca Tavazza*, Kamal Choudhary, Brian DeCost

Materials Science and Engineering Division, National Institute of Standards and Technology, Gaithersburg, MD, 20899, USA

* = Corresponding author (ftavazza@nist.gov)

## ABSTRACT

The development of large databases of material properties, together with the availability of powerful computers, has allowed machine learning (ML) modeling to become a widely used tool for predicting material performances. While confidence intervals are commonly reported for such ML models, prediction intervals, i.e., the uncertainty on each prediction, are not as frequently available. Here, we investigate three easy-to-implement approaches to determine such individual uncertainty, comparing them across ten ML quantities spanning energetics, mechanical, electronic, optical, and spectral properties. Specifically, we focused on the Quantile approach, the direct machine learning of the prediction intervals and Ensemble methods.

# 1. INTRODUCTION

In the last few years, machine learning (ML) modeling has become consistently successful across a wide variety of materials science tasks. While Material Science is still data-poor compared to the other Artificial Intelligence (AI) areas, the fact that modern instruments and powerful computational facilities have been able to produce large sets of consistent data is one of the main reasons behind such recent successes. In the area of atomistic simulations, a variety of classification and regression models have been developed for energetic[1-3], elastic[4-6], electronic[7-9], and optical properties[10], among others. JARVIS-ML[3] is an example of a publicly available repository of ML models for the prediction of a variety of properties for ideal, crystalline materials. All models in JARVIS-ML have been trained and tested on density functional theory (DFT) data contained in the JARVIS-DFT database[11, 12]. As the databases are constantly being populated, the correspondent ML models are periodically retrained, as more data has become available.

While significant effort is constantly being invested in developing new and more efficient algorithms, relatively little attention is spent in evaluating their uncertainty, beyond assessing the average ability of the model. This is even more surprising considering that ML methods are intrinsically statistical in nature. While there is a fair amount of discussion on how to evaluate uncertainty in ML/materials research[13-15], on how to estimate prediction intervals[16-21], and on how to collect data to improve ML models[22], the uncertainty on the individual predictions is often not rigorously evaluated or reported in the majority of material science ML-related papers. Currently, most such papers exclusively provide uncertainty evaluation on the average ability of ML model(s), providing quantities like the mean average error (MAE), the mean square error (MSE) or the root mean square error (RMSE). This approach fails to address the uncertainty/confidence of predictions for individual instances, i.e., how trustworthy the specific data is, which becomes important when ML-predicted data is used in other models or to aid interpretation of experimental results. Moreover, because the MSE-type stats make assumptions about the future data that we know we can't satisfy (independent, identically distributed data) in materials discovery and design settings, only providing these quantities as quality evaluators may be deceiving about the real capability of the ML model. The reason behind ML uncertainty quantification (UQ) focusing mostly on population variables is that determining prediction intervals, i.e., the uncertainty on each specific prediction, usually requires an additional computational effort beyond the training of the model. Gaussian Processes[23] are an exception to this rule, as the individual uncertainty is automatically determined when the model is fit. However, they have other limitations that often prevent them from being the approach of choice. Obviously, no amount of UQ can make up/reveal bad quality in the training data as systematic errors in the generation of the data set will propagate through a machine learning model. For example, standard DFT is known to underestimate electronic bandgaps, so any ML model trained on such data will also predict underestimated bandgaps.

In this work we use three easy-to-implement approaches to determine prediction intervals for ten machine learned physical properties. Specifically, we compare results from the Quantile approach, the direct machine learning of the prediction intervals and using a straightforward implementation of the Ensemble method. In a previous work[24], we had compared Quantile, direct ML and Gaussian processes for ML models trained and tested on DFT data available in the NIST JARVIS-DFT database[11, 12] in 2020. As the database is continuously increasing, the amount of data currently available is significantly larger than in 2020. Therefore, we have re-trained all the models for this work, as well as added the Ensemble approach to the study. We also added the comparison between Quantile results for symmetric and asymmetric implementation. However, we did drop investigating Gaussian Processes, for time constraints and because our classical force-field inspired descriptors (CFID)[3] make fitting Gaussian Processes more time-consuming than using the other approaches. As in the previous work, all investigated methods are applied to energetic, mechanical, electronic, optical, and spectral properties, to probe if/how much the findings are dependent on the physical property under examination.

## 2. METHOD

All ML models discussed in this paper are regression models as implemented in LightGBM[25] for Gradient Boosting Decision Tree (GBDT). We use JARVIS-ML based classical force-field inspired descriptors (CFID)[3]. The CFID descriptors provide a set of structural and chemical features (1557 for each material) and CFID based models have been successfully used to develop more than 25 high-accurate ML property prediction models[3] .

All the ML models in this work are trained and validated on the DFT data contained in the JARVIS-DFT database[11, 12] (8-18-2021version, https://figshare.com/articles/dataset/jdft_3d-7-7-2018_json/6815699?file=29205201). For each property, this allows for a consistent comparison between methodologies. The dataset size depends on the property under examination, as computing some properties is much more involved than computing others. The largest datasets we used were for formation energy and OPT-band gap, i.e., band gap computed using the OPTB88vdW functional (OPT) and consisted of 55723 materials. This is about 50% larger than the dataset used, for the same properties, in our previous work[24]. The smallest dataset was for exfoliation energies, counting only 813 datapoints. Table S1 gives the dataset size for each property investigated in this work.

When dealing with prediction intervals, an important concept to keep in mind is that of in-bounds count. This is a direct measure of how often the exact value falls inside the prediction interval and, therefore, it gives the prediction interval its meaning. For Gaussian distributed errors, an in-bounds count of 68 % takes the significance of 1 standard deviation. For each method under examination and each physical quantity, we used the validation set to determine its in-bound counts. When comparing prediction intervals methodologies, it's crucial that they have comparable in-bound counts as, otherwise, the prediction intervals have different meaning. In this work, we aimed at prediction intervals covering 68 % of the population. While we tried to fit ML models with such an in-bound percentage, that wasn't always possible. In such cases, we resorted to rescale the size of the prediction intervals so that the sought-after coverage was reached. Such procedure is described in more detail in our previous paper, and the rescaling factor used for each approach and each physical quantity are given in the SI.

The first methodology employed in this work estimates the prediction intervals using the Quantile approach, i.e., through the evaluation of an upper and a lower prediction bound for each predicted data. This is possible because the quantile loss function penalizes low and high errors unequally, therefore allowing to optimize by percentiles. More specifically, what percentile the quantile loss function (*quantile*) optimizes for depends on the choice of the quantile parameter $\alpha$[13]:

$$Quantile\ loss = \begin{cases} \alpha * \mid y_i - y_i^P \mid & if\ (y_i - y_i^P) \geq 0 \\ (\alpha - 1) * \left| y_i - y_i^P \right| & if\ (y_i - y_i^P) < 0 \end{cases} \qquad (1)$$

where $\alpha$ is between 0 and 1. Therefore, to determine the prediction bounds, for each property, we need to independently optimize 3 ML models, one for $\alpha_{upper}$, one for $\alpha_{lower}$ and one for $\alpha_{mid}$, where $\alpha_{mid} = 0.5$ (i.e., the median). In the rest of the paper, we will refer to these models as UPPER, LOWER and MID, respectively. More details on this approach are discussed in our previous paper and references within, including the fact that, as done in this work, modelling the MID quantities (referred to as the "predicted value" in the rest of the paper) using the least-square loss function (*ls*) instead of quantile with $\alpha_{mid} = 0.5$ generates models with smaller MAE, as it optimizes for the mean instead of for the median. In this approach, prediction intervals are given by:

$$PredInt(y_i^P) = \left| y_i^{Upper} - y_i^{Lower} \right| \qquad (2)$$

where $y_i^{Upper}$ ($y_i^{Lower}$) are the values predicted for $y_i$ using $\alpha_{upper}$ ($\alpha_{lower}$) and $y_i^P$ is the MID quantity (the predicted value of the property). As the LOWER, MID and UPPER values are determined independently, generally, the prediction interval is asymmetric with respect to the MID value. Furthermore, it may occasionally occur that the MID value doesn't fall in-between the UPPER and the LOWER ones. Fig. 1a) illustrates such possibilities, while Fig.1 b) shows how the same prediction intervals behave when symmetrized around the MID value. Such symmetrization is useful for plotting purposes and to be able to directly compare Quantile-given errors to prediction intervals obtained using the other methodologies. Therefore, in the rest of the paper we will also define a symmetrical (*SYM) quantile error* as

$$err_{Quantile}^{SYM} = PredInt(y_i^P)/2 \qquad (3)$$

where $PredInt(y_i^P)$ is given by (2). Both the symmetric and the asymmetric quantile prediction intervals will be discussed in this work and compared to each other.

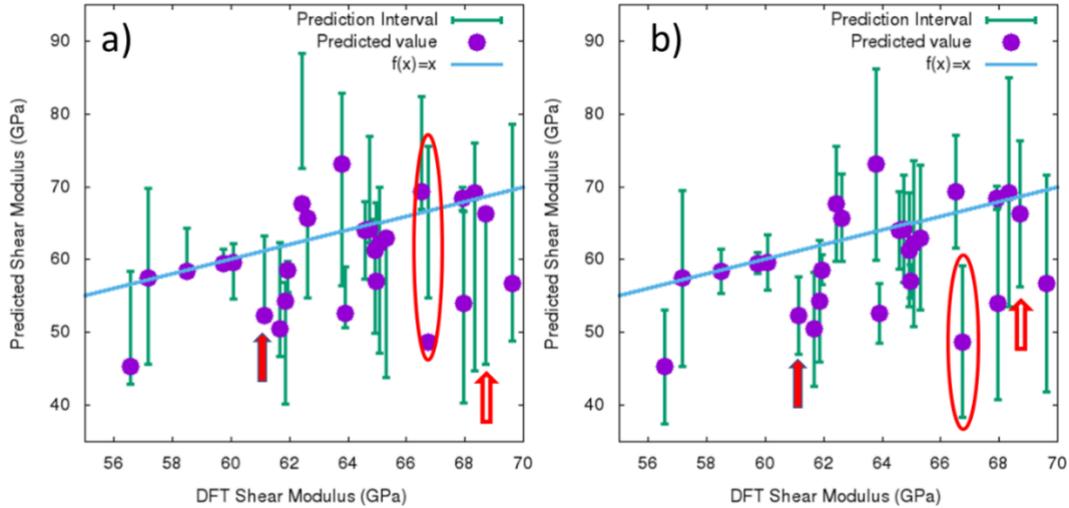

*Figure 1* Quantile prediction intervals for shear modulus (enlargement). The upper and lower predictions determine the size of the prediction interval while the "mid" prediction determines the predicted value $y_i^P$. The y=x line shows where the predicted value should be to match the DFT data. a) Asymmetric realization of the prediction intervals. The elliptical enclosure shows an example of predicted value outside the prediction interval. b) Same data but with prediction intervals symmetrized around the MID value ($err_{quantile} = PredInt(y_i^P)/2$). By construction, no case where the MID value is outside the prediction interval is now possible. The solid arrow points out a case where the prediction interval includes the DFT value (i.e., crosses the y=x line) in the asymmetric case but not in the symmetric one. The open arrow marks an opposite occurrence (DFT value included in the prediction intervals in the symmetric case, but not in the asymmetric one).

As a standard practice, when investigating the Quantile approach, we use a 90 % to 10 % train-test split[3, 6, 9]. The 10 % independent test set is never used in the hyperparameter optimization or model training, so that when the model is evaluated on it, we obtain unbiased performance metrics. In principle, the choice of $\alpha_{upper} = 0.84$ and $\alpha_{lower} = 0.14$ should lead to coverage of 68 % of the population. However, that would only occur for perfect fits. Therefore, it is the direct count of how often the reference value falls within the UPPER-LOWER interval (i.e., the in-bound counts for the Quantile approach) that provides the correct meaning of the uncertainty.

The second approach we use to determine prediction intervals uses a machine learning model to predict the errors directly. This means that it requires the fitting of two ML models for each physical quantity under examination: one to predict the actual value of the property ("base" model) and one to determine the prediction intervals ("error" model). Consequently, training the models requires splitting the data in three groups: one to fit the base model, one to fit the error model, which can also be used to validate the base model, and the last one to validate the error model. Obviously, the third set can be used to validate the base model as well.  Requiring splitting the data into three sets may be a substantial disadvantage when the amount of available data is small to begin with.  We chose to work with a 45-45-10 split, so that the number of samples in the third set of data matches the number of data used to validate the other methodologies. This makes the comparisons between approaches more consistent. In the rest of the paper this approach will be referred to as "3split", because of its requirement of splitting the available data in three groups. Specifically, we defined as "error" the absolute value of the difference between exact and predicted value. An important advantage of this second approach is that it doesn't require a specific loss function, which means that it can be used with any regression model. However, to reduce the difference between approaches to a minimum, we chose to use the same GBDT algorithm, with *ls* loss function as used to predict MID in the Quantile approach.

The third approach we investigated uses Ensemble learning [25-26], which utilizes multiple models and different random initializations to improve predictability. Because of the multiple predictions, statistical distributions of the outputs are generated. Combining such results into a Gaussian distribution, confidence intervals are obtained through variance evaluation. Such a multi-model strategy allows the evaluation of aleatory uncertainty when sufficient training data are provided. For areas without sufficient data, the predicted mean and variance will not be accurate, but the expectation is that a very large variance should be estimated, clearly indicating un-trustable predictions. In this work, for each physical quantity, we put aside 10 % of the data to compute the ensemble average on and used the remaining 90 % to train and test the various models (90-10 split). To increase difference between models, not only different random initializations were used, but also different hyperparameters. Specifically, for each model, the random number determining how the 90-10 split occurs was changed, as well as the random number controlling which hyperparameters to explore in a random search and the ranges explored in the hyperparameter search. For each property, we trained ten different models, than applied them to the 10 % of the data that was set aside. For each datapoint of that set, we computed the mean and variance, from which we determined the prediction interval as twice the standard deviation. Once again, because we are averaging only a finite number of sets and the ML models are not perfect, it is the in-bound counts, evaluated on the final set of mean values, that provides the true meaning of the prediction intervals.

## 3. RESULTS and DISCUSSION

Figure 2 shows a representative example of our findings. <mark>Only the validation data are plotted (10 % of the initial dataset, kept separate from the training data at all times).</mark> Specifically, it compares prediction intervals determined with the three approaches under examination, in the case of formation energy. As expected for reasonable models, in all three cases the smallest prediction intervals (blue dots in the colored version) are found near the diagonal, as there it is where well predicted data lie. The further from the diagonal, the larger the prediction interval should be, and that is what we found in all cases. However, the dispersion in the prediction intervals values appears to be different between models, being largest for those determined using Quantile and smallest for the Ensemble ones. As pointed out in the Method section, prediction intervals are given meaning by their corresponding in-bound count, and those are not identical between the three models. This makes comparing MAEs for the error models not as straightforward as it is comparing them for the base models. MAEs for base and error models, as well as in-bound count, are given for each approach in the inset of the corresponding figure panel. In the example under examination, all three approaches produce relatively high in-bound counts, but only Quantile is close to the desired value of 68 % (one standard deviation).

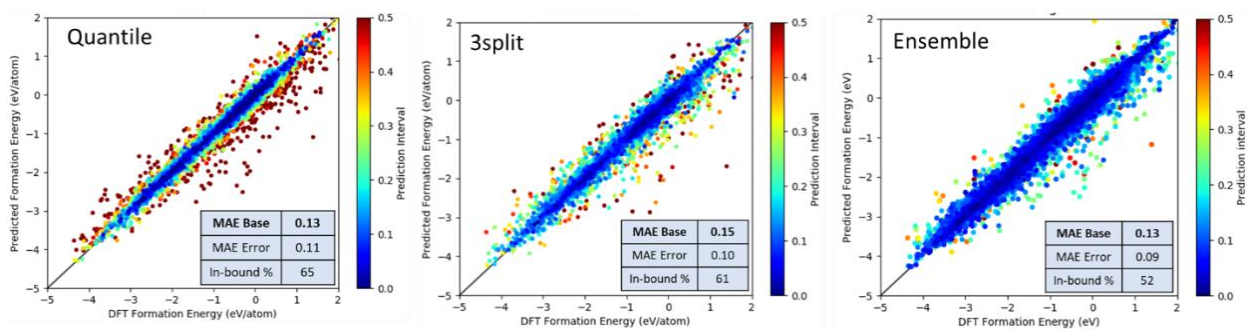

*Figure 2 Prediction Intervals for formation energy, computed using the Quantile approach (a), machine learning the error directly (3split) (b) and utilizing ensemble learning (c), where the average and standard deviation is taken over 10 independent runs. In all cases, only validation data are plotted.*

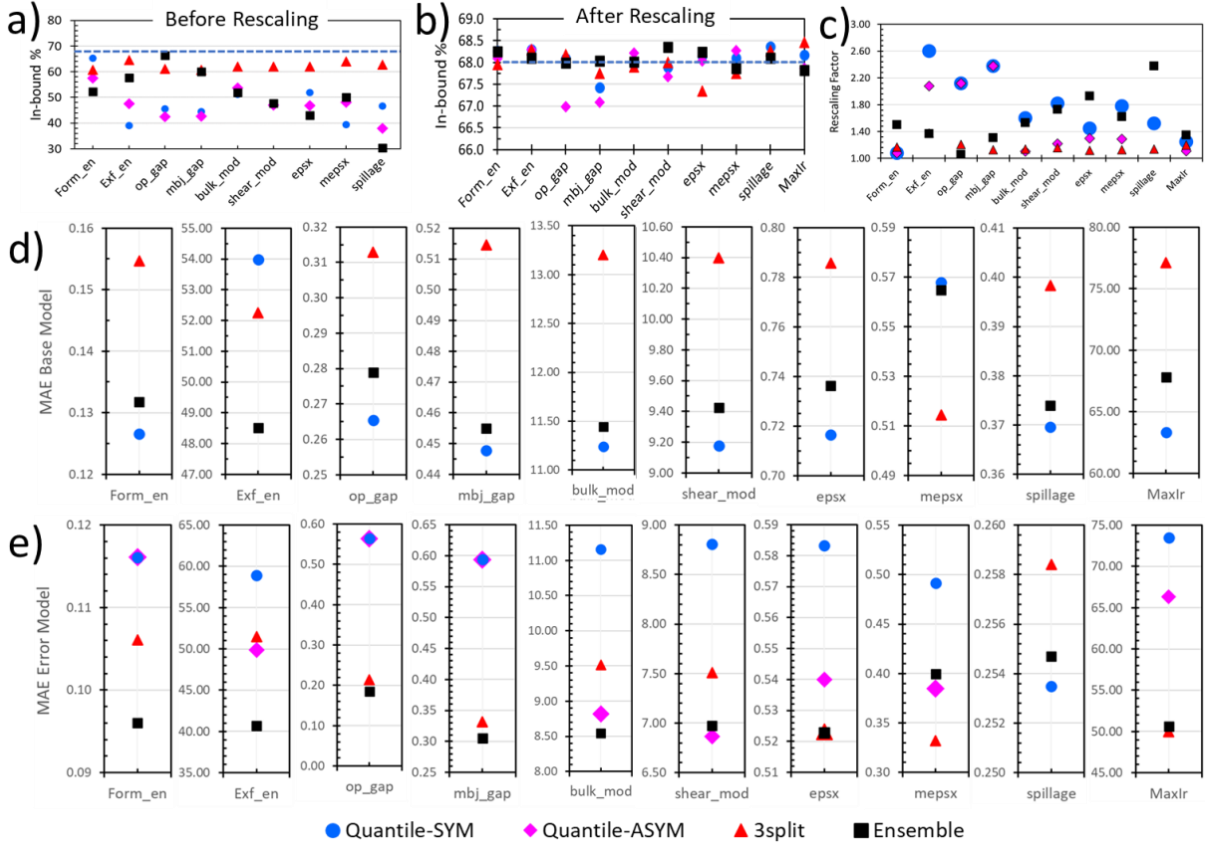

*Figure 3 In-bound counts and population variables for the 4 error models investigated in this work and all physical properties. In-bound counts before (a) and after (b) rescaling. The dashed line indicates the 68 % inbound counts that would give easy interpretability to the prediction intervals. Rescaling factor in c). Base model MAE results in d) and MAE for rescaled error models in e). Each MAE is in the units of the corresponding property (bulk_mod, shear_mod = bulk and shear modulus, respectively (GPa), op_gap, mbj_gap = bandgap using OPTB88vdW, MBJ data, respectively (eV), spillage: no units, epsx(mepsx) = refractive index along x using OPTB88vdW, MBJ data, respectively (no-units), Max_ir = maximum infrared frequency (cm-1).*

To generalize our findings, in Fig. 3 we display our results for the <mark>data in the validation sets</mark> for all methods/physical quantities under examination. Because of its significance, we start the discussion focusing on the in-bound counts. Fig.3 a) displays our findings for the lowest MAE models. While the quantile in-bound count is the highest for formation energy, as seen in Fig. 2, across the board it is the 3split approach to, more commonly, produce the count closest to the wished-for 68 %. The results are more property-dependent for the other approaches, as, for instance, ensemble performs very well for electronic quantities (energy gap, where OPT and MBJ refer to two slightly different DFT approaches) but not as well for spillage (topological property) and practically identically to Quantile for mechanical properties (bulk and shear modulus). No systematic, major difference is found between symmetrical and asymmetrical realization of Quantile.

As an accurate comparison of MAEs for the four methodologies is only possible for very similar inbounds counts, we rescaled the prediction intervals, as to obtain an inbounds percentage of 68 % (Fig. 3b). Such a rescaling was accomplished by multiplying all validation prediction intervals, for each physical property and error-modeling approach, by a fitted factor, chosen so that the inbounds percentage is 68 % or as close to it as possible. Such factors are approach- and property-dependent and are shown in Fig. 3c). While the base model MAEs (Fig. 3d) are not affected by the rescaling, the error model MAEs are then recomputed using the scaled, validation sample values and are displayed in Fig. 3 e). As expected, because for the results in Fig. 3a), 3split models require very little rescaling (factors very close to 1), while scaling factors for Ensemble and Quantile in its symmetrical realization vary depending on the physical property and can be as large as 2.6. However, with a few exceptions, scaling factors for Quantile in its asymmetrical realization are lower than those for symmetrical Quantile and tend to be very close to those for 3split. This isn't surprising, as Quantile is an intrinsically asymmetric methodology, and, therefore, it makes sense that it behaves better when used in such a form.

Focusing on the base model results (Fig. 3d), we note that the MAEs are fairly comparable between Quantile and Ensemble, while in most cases 3split-MAE is slightly higher. This is likely because of the smaller dataset available for training the base model, because of the extra data split required with this approach. The rescaled error models (Fig. 3e) show Ensemble to have the lowest MAE in most cases, followed by 3split. Asymmetric quantile's results are not far behind, while symmetric quantile is almost always having the highest MAE. These results agree with our previous work[24] as, there too, 3split was found to behave better, as error model, than Quantile. Ensemble and asymmetrical Quantile were not examined in that work.

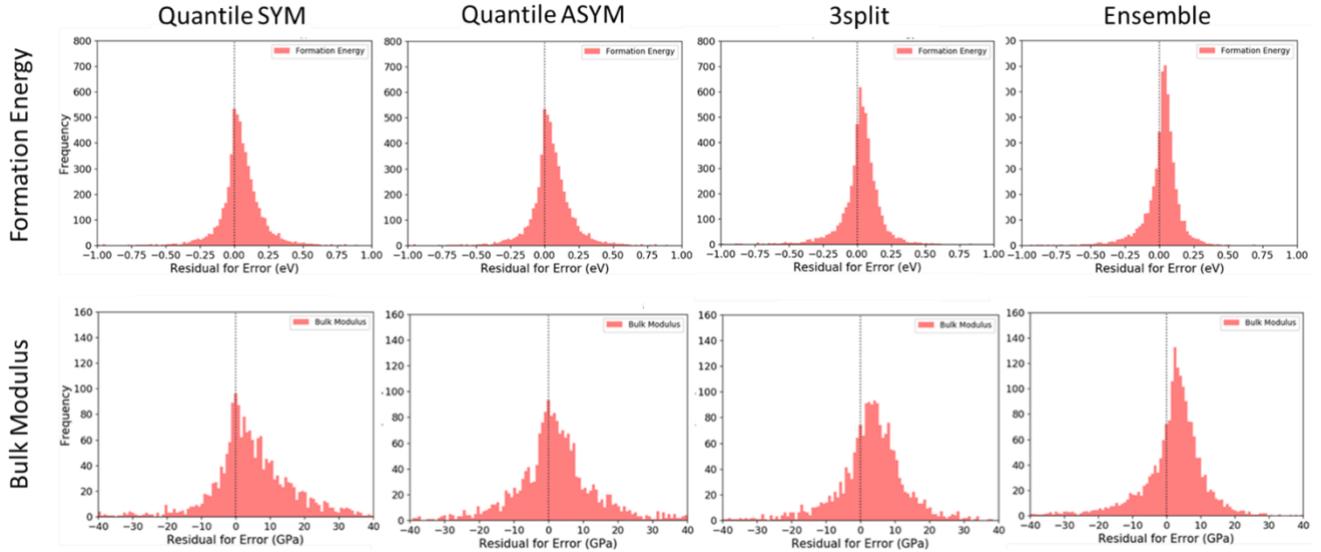

*Figure 4 : Histograms for error model residual (predicted error – exact error). Top row: formation energy, bottom row: bulk modulus. Dependence on the modeling approach is evident here, as the height, range and symmetry of the residual distribution vary among error models.*

MAE provides important information on the quality of a model, but not a complete one, as it is an averaged quantity. Investigating the distribution of residuals, for each approach, gives a more detailed picture of similarities and differences between methodologies. Such residuals are shown in Fig. 4 for the error models of Formation Energy (top row) and Bulk Modulus (bottom row). For both quantities, but especially Bulk Modulus, the dispersion of the error distribution does depend on the modeling approach. Ensemble produces the best results, as, for both quantities, its error-residual histogram is the most symmetric, is centered very close to zero, is the tallest and the one with smallest full-width-half-maximum. 3split and asymmetrical Quantile residual distributions are similar to each other, with 3split being slightly taller and narrower. Both renderings of Quantile have their highest count at zero, but symmetrical Quantile tend to overestimate the error more than its asymmetrical counterpart (longer tail towards positive residuals). Ensemble and 3split also tend to overestimate the error, as their peak is centered to the right of the zero.

## 4. CONCLUSIONS

In this work, we compare three different approaches to determine the uncertainty on individual machine learning predictions (prediction intervals). Specifically, we probe the Quantile loss function approach in its symmetrical and asymmetrical rendering, machine learning the error directly, where the error is defined either as the absolute difference (3split) and using Ensemble learning. All approaches are applied to the modeling of ten physical properties, ranging from energetic-related ones to elastic, optical, electronic, and topological. Determining the most effective way to evaluate prediction intervals, i.e., the error of each individual prediction, is necessary because quantities like MAE only evaluate ML models in a statistical manner, while users of such models need to know how reliable the specific predictions are, as that is what they most likely will use. We identify each approach advantages and disadvantages. Using the Quantile approach requires fitting three models and, in general, gives better prediction intervals when used in its asymmetric form. Machine learning the error directly has the enormous advantage of allowing the use of any loss function. However, it requires splitting the dataset in three parts, which could be a problem if the dataset is small to begin with. Ensemble learning is the easiest to implement and provides the prediction interval estimate without requiring any extra simulation, as the prediction intervals are evaluated through variance evaluation of the statistical distributions of the outputs. In-bound count is particularly important in this approach, as too small a count indicates that the models used to determine the base predictions are too close to each other to be effective in such a methodology.

All data for training and testing were taken from the publicly available JARVIS-DFT database, and the codes developed for computing the prediction intervals are available through JARVIS-Tools github (https://github.com/usnistgov/jarvis).

**Data Availability Statement:** The data from the figures are available on reasonable request from the corresponding author. Data used to train all the ML models are available on Figshare (https://figshare.com/search?q=:keyword:%20%22JARVIS%22) and examples of some of the codes used to evaluate the prediction intervals are currently available on GitHub (https://github.com/usnistgov/jarvis/tree/master/jarvis/ai/uncertainty) or will be uploaded soon. All the codes are available on reasonable request from the corresponding author.

**Acknowledgments:** Commercial equipment, instruments, software, and other materials are identified in this paper to specify the computational procedure adequately. Such identification is not intended to imply recommendation or endorsement by the National Institute of Standards and Technology or the United States government, nor is it intended to imply that the materials or equipment identified are necessarily the best available for the purpose.

**Conflicts of Interest:** The authors declare no conflict of interest. The funders had no role in the design of the study; in the collection, analyses, or interpretation of data; in the writing of the manuscript; or in the decision to publish the results.

## Supplemental Information

| Dataset size | Form. Energy | Exf. Energy | OPT gap | MBJ gap | Bulk modulus | Shear modulus | OPT epsx | MBJ epsx | spillage | Max lr |
|---|---|---|---|---|---|---|---|---|---|---|
| Number of datapoints | 55723 | 813 | 55723 | 18050 | 19109 | 18498 | 26391 | 11372 | 11372 | 3322 |

Table S1 Dataset size for each of the investigated properties